\documentclass[twocolumn,showpacs,preprintnumbers,amsmath,amssymb,aps]{revtex4-2}
\usepackage{graphicx}
\usepackage{dcolumn}
\usepackage{bm}
\usepackage[colorlinks=true]{hyperref}
\usepackage{epstopdf}
\usepackage[T1]{fontenc} 
\usepackage{lmodern}
\usepackage{braket}
\usepackage{ulem}
\usepackage{color}

\definecolor{color}{rgb}{0.11,0.45,0.02}

\begin{document}

\title{Scaling laws of electron and hole spin relaxation in indirect band gap (In,Al)As/AlAs quantum dots}

\author{T.~S.~Shamirzaev$^{1}$, D.~R.~Yakovlev$^{2,3}$, D.~S.~Smirnov$^{3}$, V.~N.~Mantsevich$^{4}$,  and M.~Bayer$^{2,5}$}

\affiliation{
\mbox{$^1$Rzhanov Institute of Semiconductor Physics, Siberian Branch of the Russian Academy of Sciences, 630090 Novosibirsk, Russia}\\
\mbox{$^2$Experimentelle Physik 2, Technische Universit\"at Dortmund, 44227 Dortmund, Germany}\\
\mbox{$^3$Ioffe Institute, Russian Academy of Sciences, 194021 St. Petersburg, Russia}\\
\mbox{$^4$Lomonosov Moscow State University, 119991, Moscow, Russia}\\
\mbox{$^5$Research Center FEMS, Technische Universit\"at Dortmund, 44227 Dortmund, Germany}}

%
%
%
%

\begin{abstract}
We investigate the electron and heavy hole spin dynamics as a function of magnetic field in ensembles of indirect band gap (In,Al)As/AlAs quantum dots (QDs) with type-I band alignment. Employing a comprehensive model that accounts for both the exciton level quartet and the magnetic-field-driven redistribution of excitons between these states via spin relaxation processes, we extract the electron ($\tau_{se}$) and heavy hole ($\tau_{sh}$) spin relaxation times as a function of magnetic field for QDs of varying sizes. Our analysis reveals that both $\tau_{se}(B)$ and $\tau_{sh}(B)$ exhibit power-law scaling behavior, yet the scaling exponents for electrons and heavy holes show markedly different evolution with QD size. For QDs with a diameter of about 9~nm, we find $\tau_{se}(B)\propto B^{-5}$ and $\tau_{sh}(B)\propto B^{-3}$. Remarkably, increasing the QD diameter to about 16~nm results in a drastic change of the scaling laws, with both $\tau_{se}(B)$ and $\tau_{sh}(B)$ following a $\propto B^{-9}$ dependence. We discuss the underlying mechanisms responsible for this size-dependent transformation of the magnetic field scaling behavior of carrier spin relaxation.

\end{abstract}

\maketitle

\section{Introduction}
\label{sec:intro}

Semiconductor quantum dots (QDs) have been broadly applied in industrial fields ranging from electronic to optical products such as solar cells, photovoltaic devices, and single-photon sources~\cite{Albaladejo-Siguan,WangQD,Somaschi}.
Theoretically, long longitudinal spin lifetimes for electrons, holes, and excitons are predicted in these structures~\cite{Khaetskii0,Khaetskii01,Khaetskii,Kroutvar,Glazov,Ivchenko60}. In this context, indirect band gap QDs, which exhibit exceptionally long exciton lifetimes extending up to several hundreds of microseconds at cryogenic temperatures~\cite{Shamirzaev84}, are particularly promising for spin-based quantum information processing~\cite{Burkard,Dyakonov}. Recently, we demonstrated that despite the long exciton lifetime, the magnetic field-induced circular polarization of the exciton photoluminescence ($P_c$) in indirect band gap (In,Al)As/AlAs QDs is not fully governed by thermodynamic parameters; specifically, the ratio of the exciton Zeeman splitting to the thermal energy does not establish a quasi-equilibrium population of the exciton states. To extract the electron ($\tau_{se}$) and heavy hole ($\tau_{sh}$) spin relaxation times, we developed a specialized technique based on simulating the long-term dynamics of $P_c$~\cite{ShamirzaevJLum288}. Using this approach, we found that $\tau_{se}$ and $\tau_{sh}$ in (In,Al)As/AlAs QDs are in the microsecond range at a magnetic field of 10~T~\cite{ShamirzaevJLum288}.

In this paper, we investigate the spin dynamics of electrons and heavy holes as a function of longitudinal magnetic field to elucidate the mechanisms governing the carrier spin relaxation in indirect band gap (In,Al)As/AlAs QDs with type-I band alignment. The magnetic field dependencies of the electron and heavy hole spin relaxation times exhibit well-defined power-law scaling, $\tau_{s} \propto B^{-n}$. Remarkably, the scaling exponents $n$ are found to depend sensitively on the QD size, revealing a rich and unexpected behavior. In small QDs (diameter $\approx$ 9–11 nm), the electron spin relaxation follows $\tau_{se} \propto B^{-5}$, in agreement with the established mechanism of phonon-assisted spin flips mediated by spin–orbit coupling. However, in large QDs with a diameter of about 16 nm, both $\tau_{se}(B)$ and $\tau_{sh}(B)$ exhibit a much steeper dependence, $\propto B^{-9}$, deviating from the $\propto B^{-5}$ behavior typically observed for carrier spin relaxation in various studies of direct and indirect band gap QDs~\cite{Kroutvar,Dunker2012,Amasha}. The hole spin relaxation shows an even richer evolution, with the scaling exponent transitioning from $B^{-3}$ in small QDs to $B^{-9}$ in the largest ones. We discuss the carrier spin relaxation mechanisms responsible for the observed exponents and their systematic variation with QD size, highlighting a transition from confinement-dominated to bulk-like behavior.

\begin{figure*}[]
\includegraphics[width=12cm]{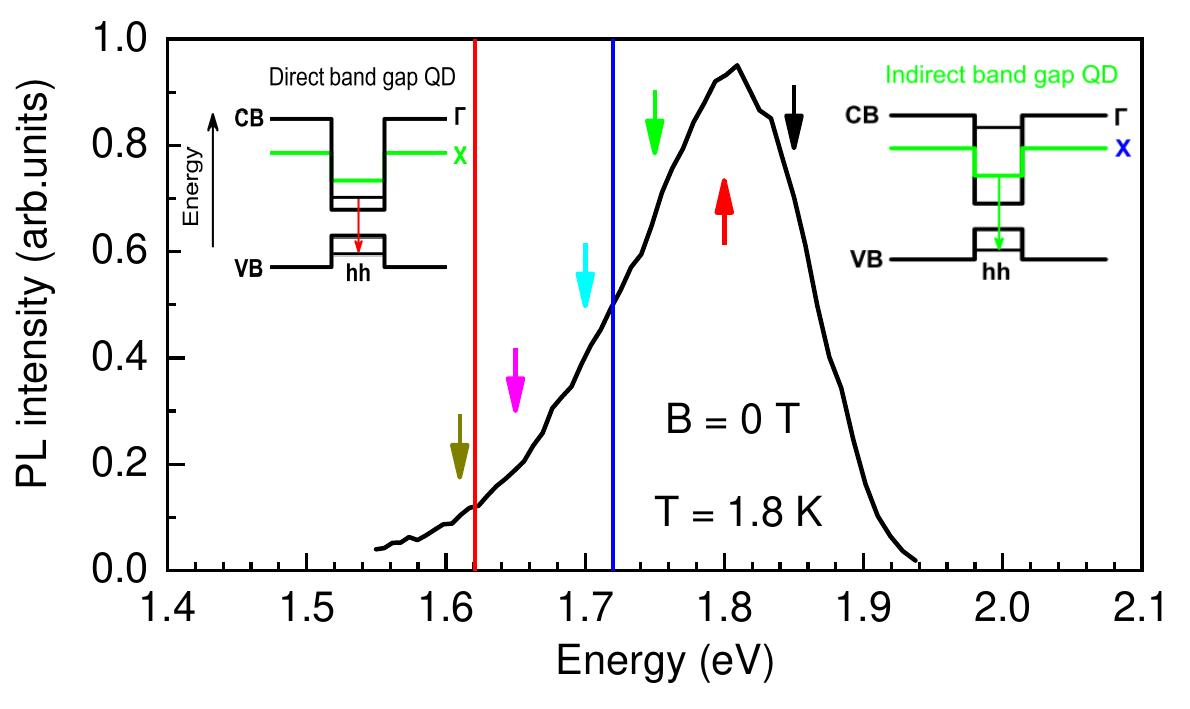}
\caption{\label{Fig1} Time-integrated PL spectrum of the (In,Al)As/AlAs QDs. Arrows mark the energies of 1.61, 1.65, 1.70, 1.75, 1.80, and 1.85 eV, where the PL and the PL polarization dynamics shown in Fig.~\ref{Fig2} were measured. Insets show schematic band diagrams of type-I (In,Al)As/AlAs QDs with direct and indirect band structure. Arrows in the insets indicate the optical transitions associated with the radiative decay of the ground state exciton.}
\end{figure*}

\section{Experimental details}
\label{sec:details}

The studied self-assembled (In,Al)As QDs, embedded in an AlAs matrix, were grown by molecular-beam epitaxy on a semi-insulating (001)-oriented GaAs substrate. The structure contains three QD sheets sandwiched between 50-nm-thick AlAs layers, which are grown on top of a 400-nm-thick GaAs buffer layer. The nominal amount of deposited InAs was approximately 2.55 monolayers. A 20-nm-thick GaAs cap layer protects the top AlAs layer against oxidation.

QD size and density were characterized by transmission electron microscopy using a JEM-4000EX system operated at an acceleration voltage of 250 keV. The density of QDs in each layer is approximately $5.2\times10^{10}$~cm$^{-2}$, with an average diameter of 13.1 nm. The QD size distribution exhibits a Gaussian profile, with half-widths corresponding to diameters of 15.8 nm (large QDs) and 8.2 nm (small QDs)~\cite{Shamirzaev78, ShamirzaevJLum288}. The determined QD composition is In$_{0.6}$Al$_{0.4}$As. Further details regarding sample growth, preparation, and atomic structure are provided in Refs.~\onlinecite{Shamirzaev78, ShamirzaevJLum288}.

The sample is mounted strain-free on a rotatable stage. It is immersed in pumped liquid helium at a temperature of $T=1.8$~K and subjected to magnetic fields up to $B=10$~T in a split-coil magnet cryostat. The magnetic field $B$ and the QD growth axis $z$ are aligned parallel to the light wave vector (Faraday geometry).

Time-integrated and time-resolved photoluminescence (PL) measurements were performed at the temperature of $T = 1.8$~K. The PL was excited by the third harmonic of a Q-switched Nd:YVO$_4$ laser (3.49 eV) with a pulse duration of 5 ns. The pulse-repetition frequency was varied from 1 to 100 kHz, and the pulse energy density was kept below 100 nJ/cm$^2$, corresponding to approximately 30\% probability of QD occupation with a single exciton~\cite{Shamirzaev84}. The PL dynamics at selected energies were detected using a GaAs photomultiplier operating in time-correlated photon-counting mode. To monitor the PL decay over a wide temporal range up to 1 ms, the time resolution of the detection system was varied between 3.2 ns and 1.5 $\mu$s.

The exciton spin dynamics were analyzed from the PL by calculating the circular polarization degree $P_c$ induced by the external magnetic field according to:
\begin{align}
P_c = \frac{I_{\sigma^+} - I_{\sigma^-}}{I_{\sigma^+} +
I_{\sigma^-}},\nonumber
\end{align}
where $I_{\sigma^+}$ and $I_{\sigma^-}$ are the intensities of the $\sigma^{+}$ and $\sigma^{-}$ polarized PL components, respectively. To determine the sign of $P_c$, we performed a control measurement on a diluted magnetic semiconductor structure with (Zn,Mn)Se/(Zn,Be)Se quantum wells, for which $P_c>0$ in Faraday geometry~\cite{Keller}.

\section{Experimental results}

\subsection{Experimental data}

Figure~\ref{Fig1} shows the time-integrated photoluminescence spectrum of the (In,Al)As/AlAs QDs measured at zero magnetic field. As we recently demonstrated, the low-energy range of the spectrum (below 1.62~eV, marked by the red line) is dominated by exciton recombination in direct-band-gap QDs, while the high-energy range (above 1.72~eV, marked by the blue line) originates from exciton recombination in indirect-band-gap QDs. Both types of QDs contribute to the intermediate spectral range~\cite{ShamirzaevJLum288}.

Our previous work also showed that the application of a longitudinal magnetic field leads to a non-uniform circular polarization of the time-integrated emission across the spectrum~\cite{ShamirzaevJLum288}. Specifically, the polarization is positive (i.e., dominated by the $\sigma^{+}$ polarized component) in the range $1.67-1.95$~eV, negative between $1.56$ and $1.67$~eV, and approaches zero for energies below $1.56$~eV.

\begin{figure}[]
\includegraphics*[width=8.5cm]{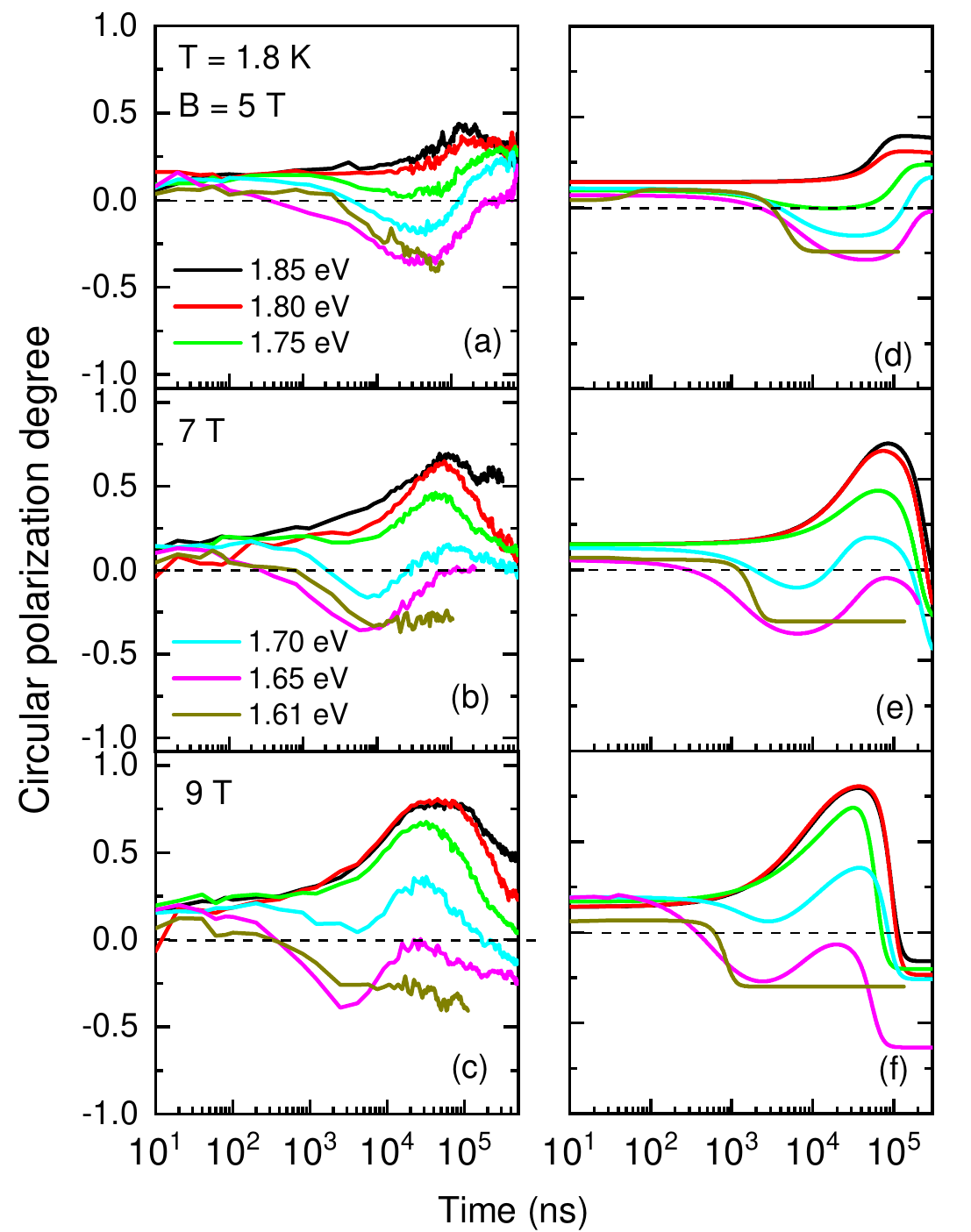}
\caption{\label{Fig2} Dynamics of the magnetic-field-induced circular polarization degree of the PL measured across the emission band (curve colors and corresponding energies are indicated by the arrows in Fig.~\ref{Fig1}) in the Faraday geometry at magnetic fields of (a) 5~T, (b) 7~T, and (c) 9~T. Panels (d), (e), and (f) show the calculated $P_{c}(t)$ dynamics at the same magnetic fields of 5, 7, and 9~T, respectively. $T = 1.8$~K.}
\end{figure}

This spectral inhomogeneity in the magnetic-field-induced polarization becomes also evident in the time-resolved measurements (for details, see Ref.~\onlinecite{ShamirzaevJLum288} and the Appendix). As shown in Fig.~\ref{Fig2}, the evolution of the circular polarization degree and its sign depend strongly on the emission energy—which reflects the QD size—and vary with the applied magnetic field. A common feature of the $P_{c}$ dynamics in strong magnetic fields is the emergence of positive circular polarization within several nanoseconds after the excitation pulse. This behavior results from rapid partial spin alignment of electrons and holes during energy relaxation in the matrix, which occurs on timescales much shorter than the spin relaxation times within the QDs~\cite{ShamirzaevJLum288,ShamirzaevJL}. The degree of this "fast" polarization is uniform across the entire spectral range.

At longer times, however, the $P_{c}$ dynamics exhibit a clear spectral dependence. In the high-energy range (corresponding to indirect-band-gap QDs), the polarization increases to a positive maximum and subsequently decreases to nearly zero. In contrast, in the low-energy range (direct-band-gap QDs), the polarization monotonically decays. The intermediate spectral range displays a more complex, non-monotonic $P_{c}$ dynamics, attributed to the superposition of contributions from both direct-like QDs (with significant $\Gamma$–X mixing~\cite{Nekrasov}) and indirect-band-gap QDs~\cite{ShamirzaevJLum288}. In this work, we focus on the spin dynamics in indirect-band-gap QDs.

\subsection{Modeling the experimental data}
\label{sec:fitting}

We recently demonstrated that the dynamics of the PL polarization degree in the studied QDs can be described within a four-level model involving the bright and dark exciton states, where spin relaxation redistributes excitons among these states. This model was originally proposed in Ref.~\onlinecite{Shamirzaev96} and later adapted for (In,Al)As/AlAs QDs in Ref.~\onlinecite{ShamirzaevJLum288}. Although the model includes several parameters, many of them can be directly determined from independent experiments, while others can be evaluated by fitting various experimental dependencies measured at $B = 10$~T. The parameters obtained in our previous studies are summarized in Table~\ref{tab:parameters} (see Appendix A). In the present work, we treat the electron ($\tau_{se}$) and heavy-hole ($\tau_{sh}$) spin relaxation times as adjustable fit parameters at fixed magnetic fields. The simulated $P_{c}(t)$ curves are shown in Figs.~\ref{Fig2}(d), \ref{Fig2}(e), and \ref{Fig2}(f) for magnetic fields of 5, 7, and 9 T, respectively. The model achieves good agreement with the experimental dynamics. The remaining discrepancies are attributed to the inhomogeneity of QD parameters—primarily the distribution of exciton lifetimes—which affects the emission at a given energy~\cite{Shamirzaev84}.

\begin{figure}[]
\includegraphics*[width=8.8cm]{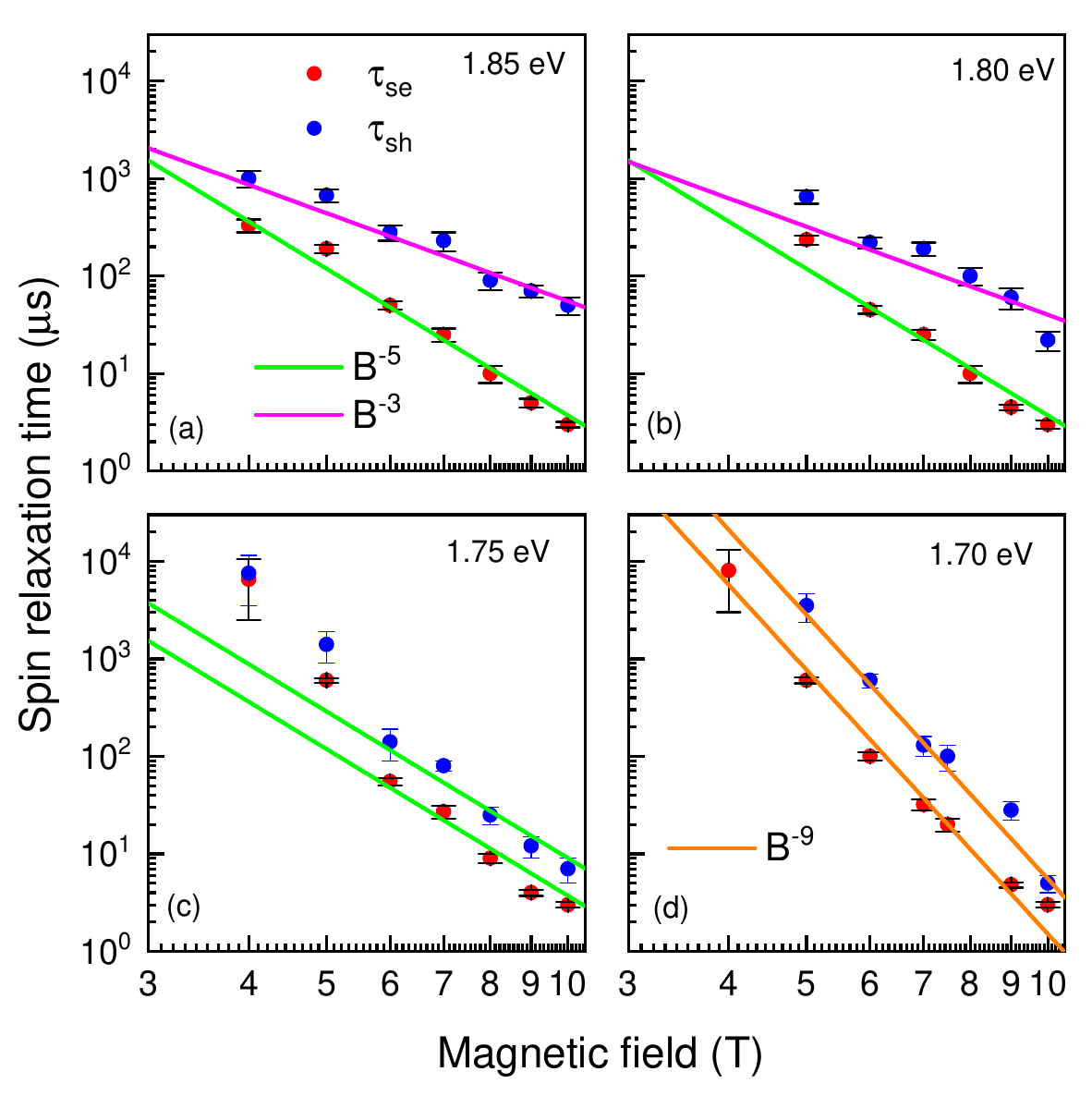}
\caption{\label{Fig3} Electron ($\tau_{se}$, red circles) and heavy-hole ($\tau_{sh}$, blue circles) spin relaxation times obtained from best fits of the experimental data at $T = 1.8$~K. The data are plotted on a double-logarithmic scale to highlight the power-law dependencies. The solid lines indicate power-law scaling with various exponents.}
\end{figure}

The best-fit values of $\tau_{se}$ and $\tau_{sh}$ for indirect-band-gap QDs at various magnetic fields are presented in Figs.~\ref{Fig3}(a)–\ref{Fig3}(d) for the exciton emission energies of 1.85, 1.80, 1.75, and 1.70 eV. These energies correspond to QD subensembles with different diameters, estimated from the correlation between the exciton transition energy and the QD size to be approximately 9, 11, 13, and 16 nm, respectively. For smaller QDs (emitting at higher energies, Figs.~\ref{Fig3}(a) and \ref{Fig3}(b)), the magnetic field dependence of the spin relaxation times follows a power law, with exponents of $B^{-5}$ for electrons and $B^{-3}$ for holes. However, as the QD size increases, deviations from these power laws appear, primarily due to a slowdown of spin relaxation at lower magnetic fields. This effect is more pronounced for the hole spin relaxation. For QDs emitting at 1.75 eV, the $\tau_{se}$ dependence deviates from $B^{-5}$ below 6 T, while the $\tau_{sh}$ dependence transitions from $B^{-3}$ to $B^{-5}$ (Fig.~\ref{Fig3}(c)). In the largest QDs (Fig.~\ref{Fig3}(d)), both $\tau_{se}$ and $\tau_{sh}$ exhibit a $B^{-9}$ dependence. Since the functional form of $\tau_{se}(B)$ and $\tau_{sh}(B)$ provides insight into the underlying spin relaxation mechanisms~\cite{Khaetskii0,Linpeng2016}, we discuss these mechanisms in the following section.

\section{Discussion}

To interpret the observed scaling of the electron and hole spin relaxation times shown in Fig.~\ref{Fig3}, we first briefly review the key theoretical results on spin relaxation mechanisms.

We begin with spin relaxation of localized electrons. Flipping an electron spin requires both energy absorption or emission and spin–orbit coupling. The necessary energy change can be provided by electron–phonon interaction. Specifically, the Zeeman splitting of electron states with $g_e = 2$ in magnetic fields of a few Tesla corresponds to the energy of acoustic phonons with wavelengths on the order of $30$~nm, which exceeds the typical QD size. In polar GaAs-like semiconductors, the piezoelectric electron–phonon coupling generally dominates over the deformation-potential mechanism~\cite{Gantmakher} (except in very high magnetic fields~\cite{Sherman2005}).

Although the spin–orbit coupling can, in principle, be incorporated directly into the phonon emission process, this mechanism is inefficient at low temperatures~\cite{Khaetskii01,Khaetskii,Woods2002}. The dominant electron spin relaxation mechanism instead involves phonon-assisted transitions between spin-mixed states arising from the Rashba and Dresselhaus spin–orbit interactions. In this case, the spin relaxation time scales as $\tau_{se} \propto B^{-5}$~\cite{Khaetskii,Woods2002}, which is in excellent agreement with the experimental data shown in Figs.~\ref{Fig3}(a) and \ref{Fig3}(b). This agreement confirms the validity of the time-resolved PL polarization model used to extract $\tau_{se}$ and $\tau_{sh}$.

At higher temperatures, where the thermal energy exceeds the Zeeman splitting, an additional power-law factor appears in the spin relaxation time due to the phonon occupation statistics~\cite{Glazov}. Other possible electron spin relaxation mechanisms include mixing of the electron spin states by nuclear spin fluctuations~\cite{Pines1957,Camenzind} and two-phonon processes~\cite{Abrahams,Khaetskii,Woods2002}, both of which may become relevant in weak magnetic fields. In the limit of vanishing Zeeman splitting, the electron spin relaxation becomes non-Markovian and is dominated by the hyperfine interaction, without involving electron--phonon coupling~\cite{Merkulov02,Loss2002,KKavokin,Glazov2015}. These considerations suggest a saturation of the electron spin relaxation time at low magnetic fields. However, in the studied (In,Al)As/AlAs indirect QDs, we observe an acceleration of this dependence, see Fig.~\ref{Fig3}(c).

Moreover, for the largest QDs (Fig.~\ref{Fig3}(d)), the dependence becomes $\tau_{se} \propto B^{-9}$. This behavior may be related to the comparable extensions of the phonon wavelength and the QD size (the magnetic length remains much larger than the QD size). In this regime, one must account for a form factor such as $\exp(-B^2/B_0^2)$, where $B_0$ is determined by the QD size (this is the magnetic field where the Zeeman energy equals the acoustic phonon energy with a wavelength of the order of the QD size). Such a factor would suppress spin relaxation and effectively yield a lower power-law exponent, contrary to the experimental observation. This effect should also become more pronounced at higher magnetic fields, which is not consistent with the trends in Fig.~\ref{Fig3}(c). To explain this, we note that the spin relaxation of electrons bound to donors in bulk semiconductors also follows $\tau_{se} \propto B^{-9}$~\cite{Linpeng2016}. This suggests strong electron localization in the largest, bulk-like QDs within the ensemble. A definitive identification of the dominant spin relaxation mechanism in this regime will require detailed microscopic modeling of the electronic structure.

Turning to the hole spin relaxation, its scaling in Fig.~\ref{Fig3} evolves from $\tau_{sh} \propto B^{-3}$ to $\propto B^{-9}$ as the QD size increases. Different forms of the spin–orbit coupling for heavy holes have been shown to result in different power laws, including $B^{-5}$, $B^{-7}$, and $B^{-9}$~\cite{Woods2004,Bulaev2005,Stano2006}. In particular, the Rashba spin–orbit coupling gives $\tau_{sh} \propto B^{-9}$~\cite{Bulaev2005}, which is expected to dominate in the strongly asymmetric QDs studied experimentally. The longer spin relaxation times observed for holes compared to electrons are also consistent with Ref.~\onlinecite{Bulaev2005}.

We note, however, that the strong variation of the hole spin relaxation scaling with exciton energy may partly arise from simplifications in the exciton PL model used to extract the spin relaxation times. In particular, the model neglects the electron–hole exchange interaction and simultaneous spin flips, which are known to be important in quantum wells~\cite{Maialle1993,Vina1999}. This issue remains open for future investigations. Nonetheless, the correct scaling of the electron spin relaxation time ($\tau_{se} \propto B^{-5}$) indicates that this quantity is reliably extracted from the experiment.

\section{Conclusions}
\label{sec:conclusions}

We have investigated the magnetic-field-induced spin dynamics of excitons in indirect-band-gap (In,Al)As/AlAs quantum dots by means of time-resolved PL spectroscopy. By modeling the experimental data within a four-level bright and dark exciton model, we have evaluated the electron ($\tau_{se}$) and heavy-hole ($\tau_{sh}$) spin relaxation times for QD subensembles with different sizes, corresponding to exciton emission energies ranging from 1.70 to 1.85~eV.

The analysis reveals distinct scaling laws of the spin relaxation times with magnetic field, which depend sensitively on the QD size. For smaller QDs (diameters $\approx$ 9–11 nm), the electron spin relaxation time follows a $\tau_{se} \propto B^{-5}$ behavior, in excellent agreement with the theoretical prediction for phonon-assisted transitions between spin-mixed states mediated by the Rashba and Dresselhaus spin–orbit coupling. The hole spin relaxation in these QDs scales as $\tau_{sh} \propto B^{-3}$.

As the QD size increases, deviations from these simple power laws emerge. In QDs with intermediate sizes ($\approx$ 13 nm), the electron spin relaxation deviates from $B^{-5}$ at low magnetic fields, while the hole relaxation exhibits a crossover from $B^{-3}$ to $B^{-5}$. In the largest QDs studied ($\approx$ 16~nm), both electron and hole spin relaxation times scale as $\tau_{se}, \tau_{sh} \propto B^{-9}$. The $B^{-9}$ dependence for electrons is reminiscent of spin relaxation of donor-bound electrons in bulk semiconductors, suggesting strong localization effects in large, bulk-like QDs. For holes, the $B^{-9}$ scaling is consistent with the dominance of the Rashba spin–orbit coupling in strongly asymmetric nanostructures.

The systematic evolution of the scaling exponents with QD size highlights the transition from confinement-dominated to bulk-like behavior and underscores the importance of size-dependent spin–orbit coupling effects. These findings provide a comprehensive picture of the spin relaxation mechanisms in indirect-band-gap QDs and demonstrate the sensitivity of the spin dynamics to quantum confinement. The extracted scaling laws serve as a benchmark for theoretical models and offer guidance for the design of QD-based spintronic devices, for which controlled spin relaxation is essential.

{\bf Acknowledgements.} The experimental activities conducted by T.S.S., including sample growth, investigation of energy level spectra and magneto-optical  properties,   as well   as   exciton   recombination   and   spin   dynamics,   were  supported  by a grant of the Russian Science Foundation (No. 22-12-00022-P). The theoretical work was supported by the Russian Science Foundation Grant No. 25-72-10031. D.S.S. and V.N.M. also acknowledge support from the Foundation for the Advancement of Theoretical Physics and Mathematics ``BASIS.''

\newpage

\appendix
\section{}

\begin{table*}[]
    \caption{Parameters of the studied (In,Al)As/AlAs QDs obtained from best fits to the experimental data at a magnetic field of 10~T~\cite{ShamirzaevJLum288}.}
  \begin{ruledtabular}
        \begin{tabular}{lccccccc}
           Parameter~$\setminus$~PL energy &  ~1.70~eV      &  ~1.75~eV  &  1.80~eV    &  1.85~eV      & \\\hline
            $g_{e}$                        &  $+2.0$       & $+2.0$     & $+2.0$      & $+2.0$        & \\
            $g_{hh}$                       &  $+2.5$       & $+2.65$    & $+2.8$      & $+2.95$       &\\
            $\tau_{r}$                     &  5.5$~\mu$s   & 12$~\mu$s  & 15$~\mu$s   & 30$~\mu$s     & \\
            $\xi$                          &  $0.85$       & $0.85$     & $0.85$      &  $0.85$       & \\
            $C_d$                          &  0.0002    &   0.0006   &   0.001     &    0.001         & \\
       \end{tabular}
  \end{ruledtabular}
    \label{tab:parameters}
  \end{table*}

The dynamics of the PL circular polarization degree, which reflects the time evolution of the carrier spin polarization in the QDs, is non-monotonic on timescales extending up to milliseconds. Both the magnitude and sign of $P_c(t)$ depend strongly on the emission energy—determined by the QD size—and vary with the applied magnetic field.

\begin{figure}[b]
\includegraphics* [width=8.5cm]{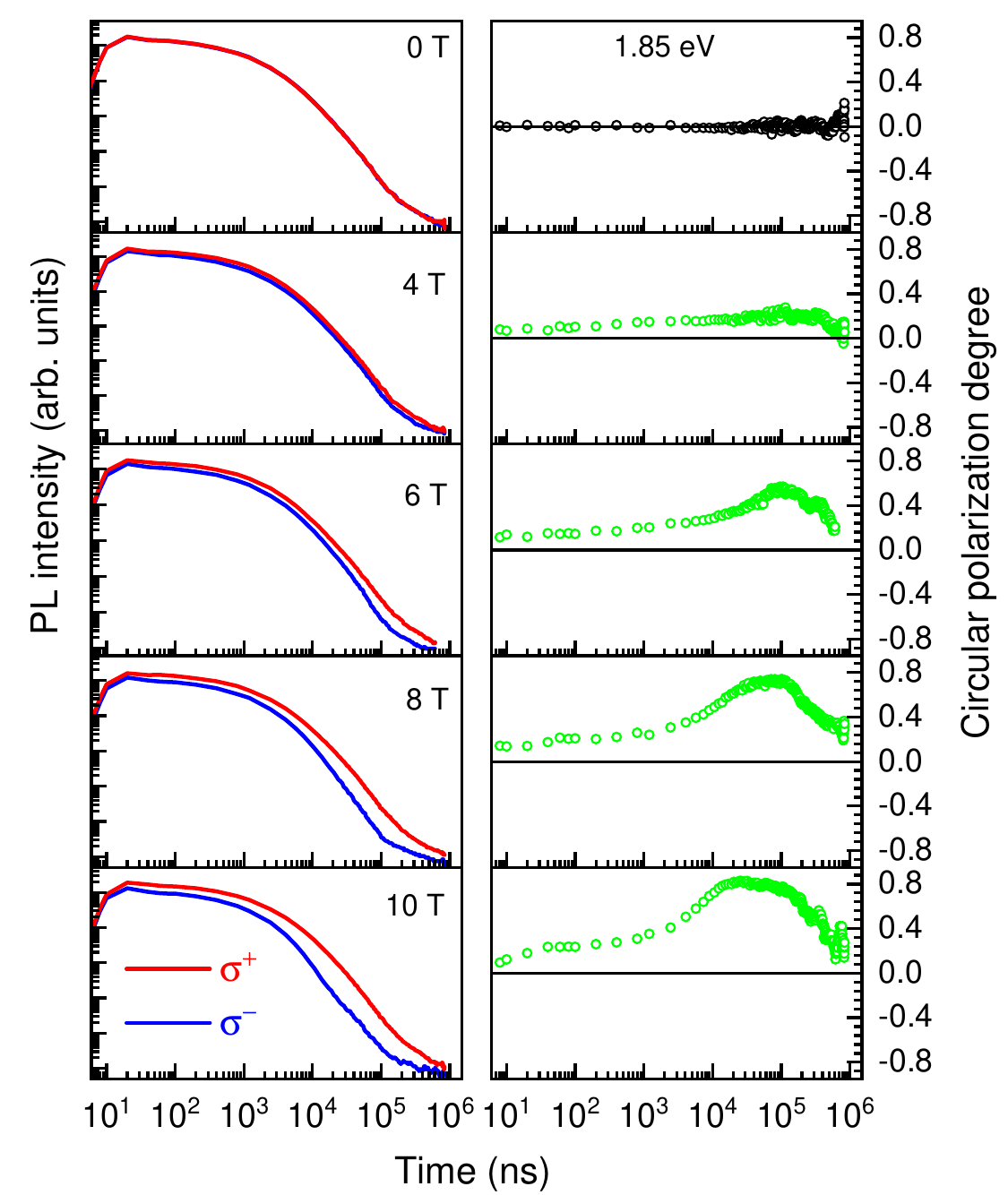}
\caption{\label{fig1A} (Left panels) PL dynamics measured in $\sigma^{+}$ (red) and $\sigma^{-}$ (blue) circular polarization as a function of magnetic field strength; (right panels) corresponding dynamics of the magnetic-field-induced PL circular polarization degree. The measurements were performed in the Faraday geometry at $T = 1.8$~K with detection at 1.85~eV.}
\end{figure}

In the high-energy spectral range (at 1.85 eV, see Fig.~\ref{fig1A}), as the longitudinal magnetic field increases from 0 to 3 T, circular polarization appears immediately after the excitation pulse and remains constant over time up to 1 ms. Upon further increasing the field strength (from 6 to 10 T), the $P_c(t)$ dynamics become non-monotonic: the polarization increases over tens of microseconds to a maximum positive value $P_c^{\text{max}}$, which depends on the field strength (reaching $+0.82$ at $B = 10$ T), and then decays to nearly zero at times exceeding a hundred microseconds.

\begin{figure}[]
\includegraphics* [width=8.0cm]{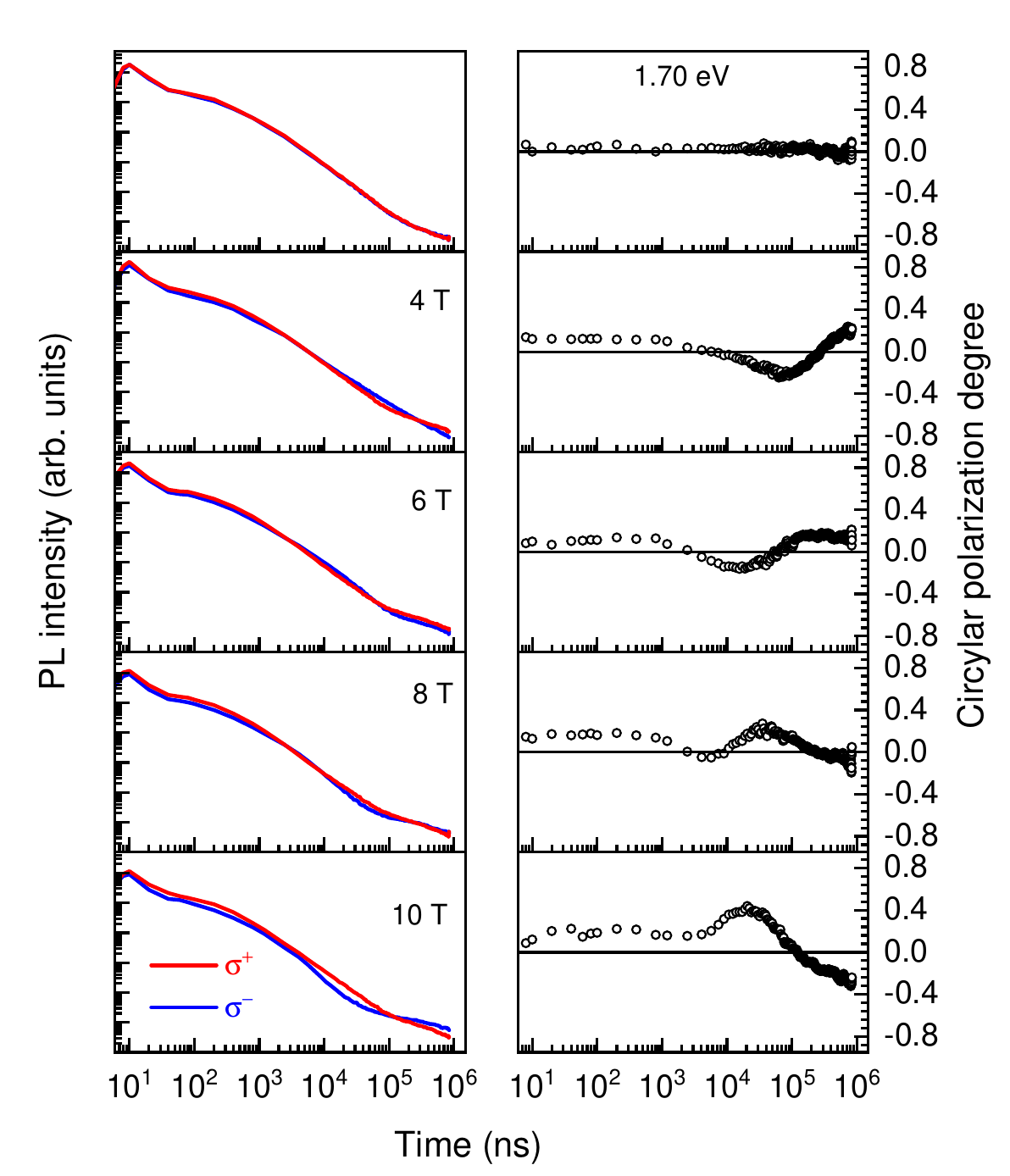}
\caption{\label{fig2A}(Left panels) PL dynamics measured in $\sigma^{+}$ (red) and $\sigma^{-}$ (blue) circular polarization as a function of magnetic field strength; (right panels) corresponding dynamics of the magnetic-field-induced PL circular polarization degree. The measurements were performed in the Faraday geometry at $T = 1.8$~K with detection at 1.70~eV.}
\end{figure}

In the low-energy spectral range (at 1.70 eV, see Fig.~\ref{fig2A}), the positive polarization begins to decay to zero and changes sign already at $B = 2$ T on timescales of about 100 $\mu$s. As the longitudinal magnetic field increases further (from 4 to 10 T), the $P_c(t)$ dynamics become increasingly complex, exhibiting non-monotonic behavior and multiple sign changes over time.

The parameters of the four-level model (involving bright and dark exciton states and their redistribution via spin relaxation), which describes the dynamics of the PL polarization degree and were determined in our previous studies~\cite{ShamirzaevJLum288}, are summarized in Tab.~\ref{tab:parameters}.


\newpage



\begin{thebibliography}{}

\bibitem{Albaladejo-Siguan} M.~Albaladejo-Siguan, E.~C.~Baird, D.~Becker-Koch, Y.~Li, A.~L.~Rogach, and Y.~Vaynzof, Stability of quantum dot solar cells: A matter of lifetime, Adv. Energy Mater. \textbf{11}, 2003457 (2021).

\bibitem{WangQD} \textit{Quantum Dot Devices}, ed. by Z.~M.~Wang  (Springer, New York, 2016).

\bibitem{Somaschi} N.~Somaschi, V.~Giesz, L.~De~Santis, J.~C.~Loredo, M.~P.~Almeida,  G.~Hornecker,  S.~L.~Portalupi,  T.~Grange,  C.~Anton, J.~Demory, C.~Gomez, I.~Sagnes, N.~D.~Lanzillotti-Kimura, A.~Lemaitre, A.~Auffeves, A.~G.~White, L.~Lanco, and P.~Senellart, Near-optimal single-photon sources in the solid state, Nat. Photonics \textbf{10}, 340 (2016).

\bibitem{Khaetskii0} A.~V.~Khaetskii and Yu.~V.~Nazarov, Spin relaxation in semiconductor quantum dots, Phys. Rev. B \textbf{61}, 12639 (2000).

\bibitem{Khaetskii01} A.~V.~Khaetskii, Spin relaxation in semiconductor mesoscopic systems, Physica E \textbf{10}, 27 (2001)

\bibitem{Khaetskii} A. V. Khaetskii, and Y. V. Nazarov, Spin-flip transitions between Zeeman sublevels in semiconductor quantum dots, Phys. Rev. B {\bf 64}, 125316 (2001).

\bibitem{Kroutvar} M.~Kroutvar, Y.~Ducommun, D.~Heiss, M.~Bichler, D.~Schuh, G.~Abstreiter, and J.~J.~Finley, Optically programmable electron spin memory using semiconductor quantum dots, Nature \textbf{432}, 81 (2004).

\bibitem{Glazov} M.~M.~ Glazov, \textit{Electron  and  Nuclear  Spin  Dynamics  in Semiconductor Nanostructures}  (Oxford University Press, Oxford, UK, 2018).

\bibitem{Ivchenko60} E.~L.~Ivchenko, Magnetic circular polarization of exciton photoluminescence,  Phys. Solid State \textbf{60}, 1514 (2018).

\bibitem{Shamirzaev84} T.~S.~Shamirzaev, J.~Debus, D.~S.~Abramkin, D.~Dunker, D.~R.~Yakovlev, D.~V.~Dmitriev, A.~K.~Gutakovskii, L.~S.~Braginsky, K.~S.~Zhuravlev, and M.~Bayer, Exciton recombination dynamics in an ensemble of (In,Al)As/AlAs quantum dots with indirect band-gap and type-I band alignment, Phys. Rev. B \textbf{84}, 155318 (2011).

\bibitem{Burkard} G.~Burkard, T.~D.~Ladd, A.~Pan, J.~M.~Nichol, and J.~R.~Petta, Semiconductor spin qubits, Rev. Mod. Phys. \textbf{95}, 025003 (2023).

\bibitem{Dyakonov} \textit{Spin Physics in Semiconductors}, ed. M.~I.~Dyakonov (Springer, Berlin, 2008).

\bibitem{ShamirzaevJLum288} T.~S.~Shamirzaev, D.~R.~Yakovlev, V.~N.~Mantsevich, D.~Kudlacik, A.~Yu.~Gornov, A.~K.~Gutakovskii, and M.~Bayer, Magnetic field induced exciton spin dynamics in indirect band gap (In,Al)As/AlAs quantum dots, J. Lumin. \textbf{288}, 121596 (2025).

\bibitem{Dunker2012} D.~Dunker, T.~S.~Shamirzaev, J.~Debus, D.~R.~Yakovlev, K.~S.~Zhuravlev, and M.~Bayer,
Spin relaxation of negatively charged excitons in (In,Al)As/AlAs
quantum dots with indirect band gap and type-I band alignment, Appl.
Phys. Lett. \textbf{101}, 142108 (2012).

\bibitem{Amasha} S. Amasha, K. MacLean,  Iuliana P. Radu, D. M. Zumb{\"u}hl,  M. A. Kastner, M. P. Hanson, and A. C. Gossard, Electrical Control of Spin Relaxation in a Quantum Dot, Phys. Rev. Lett. \textbf{100}, 046803  (2008).

\bibitem{Shamirzaev78} T.~S.~Shamirzaev, A.~V.~Nenashev, A.~K.~Gutakovskii, A.~K.~Kalagin, K.~S.~Zhuravlev, M.~Larsson, P.~O.~Holtz,
Atomic and energy structure of InAs/AlAs quantum dots, Phys. Rev. B
\textbf{78}, 085323 (2008).

\bibitem{Keller} D.~Keller, D.~R.~Yakovlev, B.~K\"onig, W.~Ossau, Th.~Gruber, A.~Waag, L.~W.~Molenkamp, and A.~V.~Scherbakov, Heating of the magnetic ion system in (Zn,Mn)Se/(Zn,Be)Se semimagnetic quantum wells by means of photoexcitation, Phys. Rev. B \textbf{65}, 035313 (2002).

\bibitem{ShamirzaevJL} T.~S.~Shamirzaev, D.~R.~Yakovlev, D.~Kudlacik, C.~Harkort, M.~A~Putyato, A.~K.~Gutakovskii, and M.~Bayer, Thin Ga(Sb,P)/GaP quantum wells with indirect band gap: Crystal structure, energy spectrum, exciton recombination and spin dynamics, J. Lumin. \textbf{277}, 120888 (2025).

\bibitem{Nekrasov} S.~V.~Nekrasov, N.~O.~Mikhailenko, M.~D.~Ragoza, T.~S.~Shamirzaev, Yu.~G.~Kusrayev, Influence of $\Gamma$-X mixing on optical orientation and alignment of excitons, Phys. Rev. B \textbf{110}, 115435 (2024).


\bibitem{Shamirzaev96}  T.~S.~Shamirzaev, J.~Rautert, D.~R.~Yakovlev, J.~Debus, A.~Yu.~Gornov, M.~M.~Glazov, E.~L.~Ivchenko, and M.~Bayer, Spin dynamics and magnetic field induced polarization of excitons in ultrathin GaAs/AlAs quantum wells with indirect band gap and type-II band alignment, Phys. Rev. B \textbf{96}, 035302 (2017).

\bibitem{Linpeng2016} X.~Linpeng, T.~Karin, M.~V.~Durnev, R.~Barbour,  M.~M.~Glazov,  E.~Ya.~Sherman,  S.~P.~Watkins,  S.~Seto, and Kai-Mei~C.~Fu, Longitudinal spin relaxation of donor-bound electrons in direct band-gap semiconductors, Phys. Rev. B \textbf{94}, 125401 (2016).

\bibitem{Gantmakher} V.~F.~Gantmakher and Y.~B.~Levinson, \textit{Carrier Scattering in Metals and Semiconductors} (North Holland, New York, 1987).

\bibitem{Sherman2005} E. Y. Sherman and D. J. Lockwood, Spin relaxation in quantum dots with random spin-orbit coupling, Phys. Rev. B
\textbf{72}, 125340 (2005).

\bibitem{Woods2002} L.~M.~Woods, T.~L.~Reinecke, and Y.~Lyanda-Geller, Spin relaxation in quantum dots, Phys. Rev. B \textbf{66}, 16318(R) (2002).

\bibitem{Pines1957} D.~Pines, J.~Bardeen and C.~P.~Slichter, Nuclear polarization and impurity-state spin Relaxation processes in silicon, Phys. Rev. \textbf{106}, 489 (1957).

\bibitem{Camenzind} L.~C.~Camenzind, L.~Yu, P.~Stano, J.~D.~Zimmerman, A.~C.~Gossard, D.~Loss and D.~M.~Zumb\"uhl, Hyperfine-phonon spin relaxation in a single-electron GaAs quantum dot, Nat. Commun. \textbf{9}, 3454 (2018).

\bibitem{Abrahams} E.~Abrahams, Donor electron spin relaxation in silicon, Phys. Rev. \textbf{107}, 491 (1957).

\bibitem{Glazov2015} M.~M.~Glazov, Spin noise of localized electrons: Interplay of hopping and hyperfine interaction, Phys. Rev. B \textbf{91}, 195301 (2015).

\bibitem{Merkulov02} I.~A.~Merkulov, A.~L.~Efros, and M.~Rosen, Electron spin relaxation by nuclei in semiconductor quantum dots, Phys. Rev. B. \textbf{65}, 205309 (2002).

\bibitem{Loss2002} A.~V.~Khaetskii, D.~Loss, and L.~Glazman, Electron Spin Decoherence in Quantum
Dots due to Interaction with Nuclei, Phys. Rev. Lett. \textbf{88}, 186802 (2002).

\bibitem{KKavokin} K.~V.~Kavokin, Spin relaxation of localized electrons in $n$-type semiconductors, Semicond. Sci. Technol. \textbf{23}, 114009 (2008).

\bibitem{Bulaev2005}  D.~V.~Bulaev and D. Loss, Spin Relaxation and Decoherence of Holes in Quantum Dots, Phys. Rev. Lett. \textbf{95}, 076805 (2005).

\bibitem{Woods2004} L.~M.~Woods, T.~L.~Reinecke, and R.~Kotlyar, Hole spin relaxation in quantum dots, Phys. Rev. B  \textbf{69}, 125330 (2004).

\bibitem{Stano2006} P.~Stano and J.~Fabian, Theory of Phonon-Induced Spin Relaxation in Laterally Coupled Quantum Dots, Phys. Rev. Lett. \textbf{96}, 186602 (2006).

\bibitem{Maialle1993} M.~Maialle, E.~de~Andrada e Silva, and L.~Sham, Exciton spin dynamics in quantum wells, Phys. Rev. B \textbf{47}, 15776 (1993).

\bibitem{Vina1999} L.~Vi{\~{n}}a, Spin relaxation in low-dimensional systems, J. Phys.: Condens. Matter \textbf{11}, 5929 (1999).


\end{thebibliography}
\end{document}